OBSERVATION OF EXCESS FLUX FOR NEGATIVE COSMIC RAY PENETRATING PARTICLES IN BUBBLE CHAMBER "SKAT" FOR MOMENTUM RANGE (30GeV/c<P<150GeV/c )

Yu.N.Bazhutov,[1] D.S.Baranov[2]

[1] State Technical University ( MADI ),  P.O. Box 169, 105077 Moscow, Russia
[2] High Temperature Institute RAS, Izhorskaya 13/19, 127412 Moscow, Russia
Bazhutov@hotmail.com  / Fax: 7-095-151-0331

ABSTRACT.   There are presented the first results of the new heavy stable cosmic ray particles search in the bubble chamber "SKAT" (450 x 160 x 90 $cm^3$), which was exposed in the neutrino beam of Serpukhov Accelerator during 1976 – 1992 years and was viewed along the horizontal direction so as the magnet field direction (MDM > 150 GeV/c). From looking over 1,270 stills (1 roll for April 23, 1979) it was selected 757 tracks of cosmic ray particles with zenith angle  - $\theta$ < 45°, track length – L > 50 cm and momentum - P>2.0GeV/c. From this events there were constructed momentum spectrums for both negative and positive vertical cosmic ray penetrating particles in the (2.0 – 126) GeV/c range and calculated their charge ratio. For positive particles the momentum spectrum has normal shape in all studied range the same as for negative particles but only for momentum range (2.0 – 32) GeV/c and charge ratio for this range is normal and the same as for cosmic muons. But for momentum - P>32GeV/c it was observed negative particles excess flux (~$10^5 cm^{-2} s^{-1} sr^{-1}$) with changed charge ratio – R = 0.62 +/- 0.18  ($\Delta$>3.5$\sigma$) for momentum range (32GeV/c<P<126GeV/c). The same anomalous results were observed from another "SKAT" 3 rolls (1717 stills in December 11-12, 1991). Here it was measured from selected near vertical 736 events for momentum range (36GeV/c<P<180GeV/c)  the charge ratio – R = 0.67 +/- 0.19  ($\Delta$>3$\sigma$) and for momentum range (3.6GeV/c<P<36GeV/c) normal charge ratio – R = 1.29 +/- 0.14.

INTRODACTION.   The first positive results on new heavy charged stable cosmic ray particles registration, received on the scintillation telescope Doch-4 in July 1999 [1-3], initiated their search on another already existed installations, which have sensitiveness to some kind of these particles parameters. The main of these parameters are following: 1) the possibility to register charged particles in near vertical direction with large acceptance and operation time ($S\Omega T>10^7 cm^2 \cdot s \cdot sr$); 2) the installation place must be on the Earth surface or small underground (< 100 m.w.e.); 3) it is desirable to use track detector for event and particle charge viewing. One of the largest Bubble Chamber "SKAT" (BC), operated from 1976 to 1992 on the Serpukhov Accelerator neutrino beam had satisfied to all these demands. Its operation on the neutrino beam could provide     a small background from Accelerator for our researches. Accumulated  ~ 4 x 2,000,000 stills inside ~ 4 x 1660 rolls during this long period were conserved until now and were ready to analysis. The large Bubble Chamber (450 x 160 x 90 $cm^3$) had been placed in greatest magnetic field (17 kG) horizontally directed so as coaxial to it view of 4 stereo photo chambers. So near vertical penetrating cosmic rays could be registered the same as near horizontal neutrino beam events. Full "SKAT" exposition is $S\Omega T \sim 4 \cdot 10^9$ $cm^2 \cdot s \cdot sr$ , that is rather more demanded one. It provides us for charge and momentum (MDM > 150 GeV/c) measurements. BC had been placed on the Earth surface, but had large magnetic iron screen (d ~ 2500 $g/cm^2$ = 25m.w.e., Fig.1).

INSTALLATION. The Bubble Chamber "SKAT" had been constructed 30 years ago [4-6] in Institute of High Energy Physics (IHEP) for operating on the Serpukhov Proton Accelerator to register neutrino interaction events. During 16 years operating (1976-1992) it had 42 runs, each of them had lasted near week continuously in different months. The intervals between every BC operation was near 10s with its sensitive time for track registration - $\tau \sim 25$ms. The BC was synchronized with accelerator beam events but cosmic rays could come in any time. So for cosmic ray events we couldn't know anything about their ionization losses. In different time BC had been filled by different liquids: 1) freon – $CF_3Br$; 2) propane – $C_3H_8$; 3) ethane – $C_2H_6$. In IPHE rolls review was realized with help of special reviewing tables with tracks scanning and computer analyzing due to special program. One of us ( Baranov D.S.) used this technique in simple variant. In MADI Bazhutov Yu.N. used another reviewing technique – Universal Measuring Microscope (UIM-21). It has highest angular resolution ($\delta \sim 1'$) and space resolution ($\Delta \sim 10 \mu m$) in both orthogonal projections at objective magnifying coefficient – k=15 to the display view on screen.

RESULTS. At the first step our research included only statistic but not individual analysis. The BC "SKAT" was used such as magnetic spectrometer for momentum spectrum constructing.
In MADI it was reviewed one roll, # 153 (1265 stills) for April 23, 1979 (3.15 – 6.25 am). From all tracks it was selected only those which started in BC ceiling and had track length more then $L_0$=50cm=76g/cm$^2$ (BC filled by liquid freon) in near vertical direction ($\theta$<45°). From the difference between start end final track angles it was calculated crookedness radius due to measured track length (L) and then particle momentum accordingly. The measurements were fulfilled only in one of 4-th corresponding rolls, in one projection but for statistic analysis it didn't input large error. As it was shown later by calibration measurements with IHEP technique the statistic momentum error was only about 15%. After final events selection with momentum – P>2.0 GeV/c it was received 757 track events ($N_+$=398; $N_-$=359). Based on these events it was constructed 3 types of differential momentum spectra (separately for positive particles, for negative ones and summarized ones; shown at fig.2). From these spectra it was constructed differential charge ratio dependence for the same momentum range (2.0GeV/c<P<126GeV/c), shown on fig.3. For positive particles the momentum spectrum has normal shape in all studied range the same as for negative particles but only for momentum range (2.0 – 32) GeV/c and charge ratio for this range is normal and the same as for cosmic muons. But for momentum - P > 32 GeV/c it was observed negative particles excess flux ($\sim 10^{-5}$cm$^{-2}$s$^{-1}$sr$^{-1}$) with changed charge ratio – R=0.62 +/- 0.18 ($\Delta$>3.5$\sigma$) for momentum range (32GeV/c<P<126GeV/c).
In IHEP it was reviewed 3 small rolls # 1511-1513 (1717 stills) for December 11-12, 1991 (8.50pm-1.55am). The selection demands was the same as in MADI, but only the minimum of track length was twice more ($L_0$=100cm=51g/cm$^2$) and BC was filled that time by liquid ethane. It was selected 736 events with momentum – P>$P_0$=3.6GeV/c ($N_+$=405; $N_-$=331). It was constructed differential momentum spectra for positive and negative penetrating particles for momentum range (36GeV/c<P<180GeV/c) with similar anomalous excess for negative ones for momentum – P>36GeV/c. From these differential momentum spectra it was received the differential charge ratio dependence (fig.4). As the result we have for momentum range (36GeV/c<P<180GeV/c) the charge ratio –R=0.67 +/- 0.19 ($\Delta$>3$\sigma$) and for momentum range (3.6GeV/c<P<36GeV/c) normal charge ratio – R = 1.29 +/- 0.14.

CONCLUSIONS. Repeated review of the roll # 153 in IHEP with special geometric reconstruction for all tracks confirmed rightness of results received in MADI. This result couldn't be connected with neutrino beam in which neutrino had momentum less then 30GeV/c in horizontal direction. Such way the authors consider that they have registered the excess flux for negative penetrating particles in cosmic rays for momentum range (30GeV/c<P<150GeV/c) at the level – J$\cong 10^{-5}$cm$^{-2}$s$^{-1}$sr$^{-1}$ for two different time duration. We understand that momentum spectrum and charge ratio for cosmic muons inside momentum range (30–150GeV/c) are well known research field with only small indications to such effect [7-10] but we are sure of our results. Now review of another "SKAT" rolls is going on. The first results indicate that not every roll has such anomalous effect.

ACKNOLEDGMENTS. We want gratefully thank for help and opportunity to review BC "SKAT" films to Drs. V.V.Ammosov, A.A.Ivanilov and for continuous support in this work to Drs. P.F.Ermolov, V.S.Murzin, L.G.Sapogin, V.P.Nosov, O.P.Gudzhoyan & V.G.Grishin.

Fig.1. Bubble Chamber "SKAT" front view.

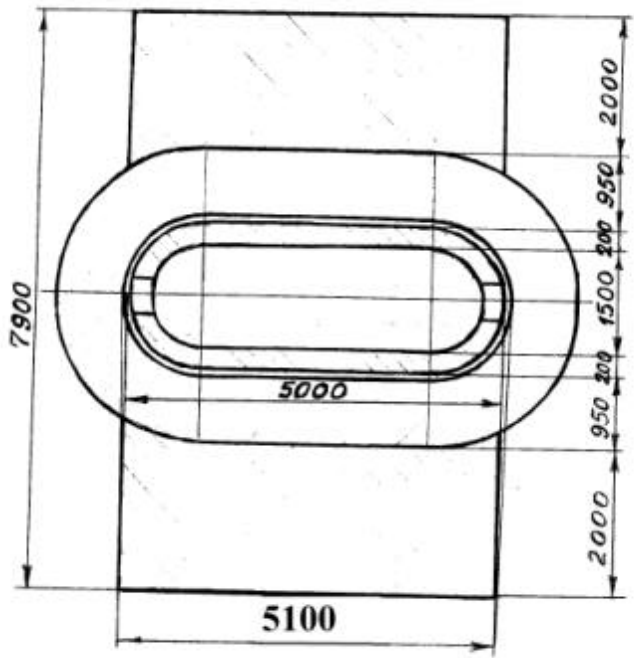

Fig.2. Differential momentum spectrum for positive, negative and all charged particles.

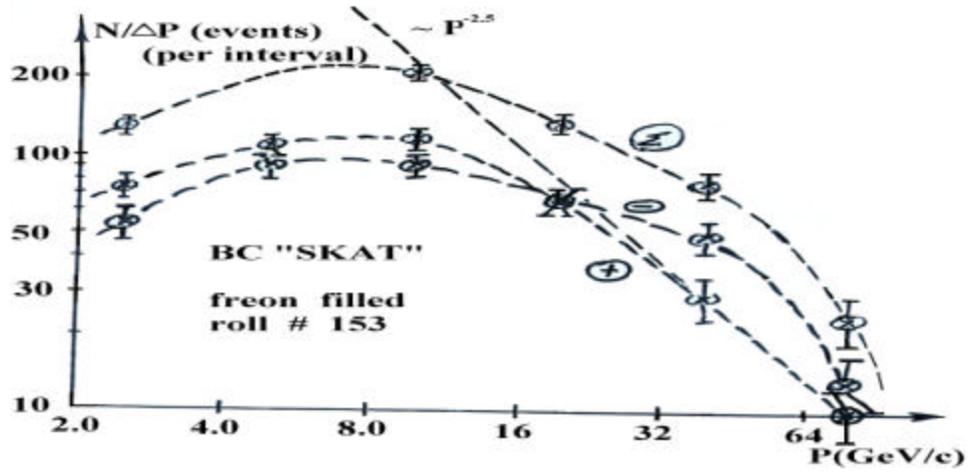

Fig.3. Charge ratio momentum dependence for events from roll # 153.

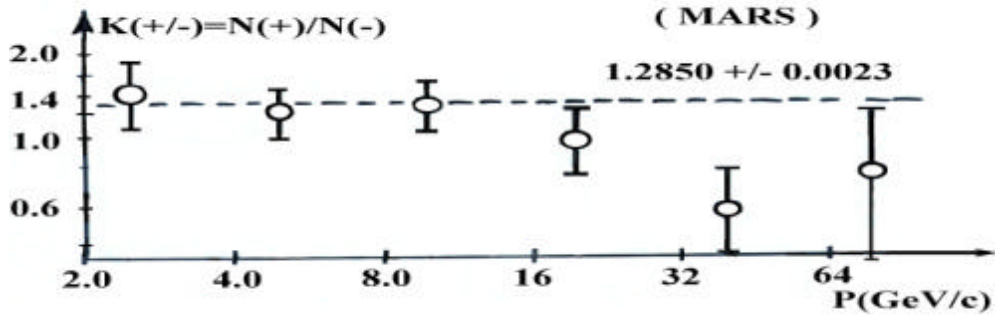

Fig.4. Charge ratio momentum dependence for events from rolls # 1511-1513.

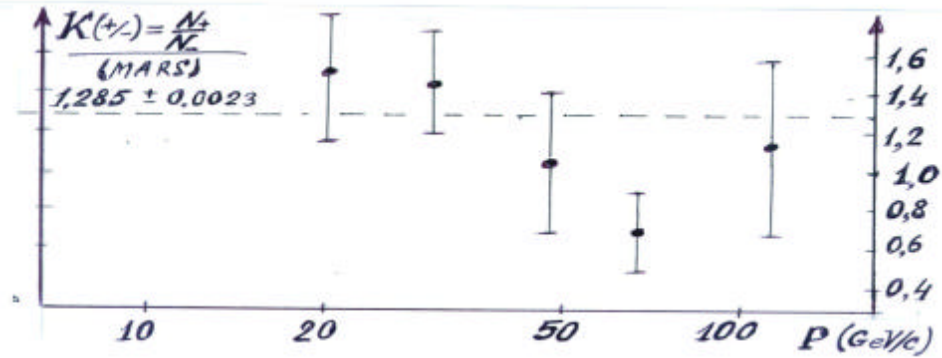